\documentclass[preprint,aps]{revtex4}
\usepackage{amsfonts}
\usepackage{amsmath}
\usepackage{amssymb}
\usepackage{array}
\usepackage{graphicx}

\begin{document} 

\title{Black-body radiation shift of atomic energy-levels:\\ The $ (Z\alpha)^2\alpha T^2/m $ correction}

\author{Wanping Zhou}
\affiliation{School of Physics and Technology, Wuhan University, Wuhan 430000 china}

\affiliation{Engineering and Technology College, Hubei University of Technology, Wuhan 430000 china}
\author{Xuesong Mei}
\affiliation{School of Physics and Technology, Wuhan University, Wuhan 430000 china}

\author{Jinjun Lu}
\affiliation{School of Physics and Technology, Wuhan University, Wuhan 430000 china}

\affiliation{School of Physics and Electronic-information Engineering, Hubei Engineering University, Xiaogan 432000, China}

\author{Haoxue Qiao \footnote{Haoxue Qiao; electronic mail: qhx@whu.edu.cn}}
\affiliation{School of Physics and Technology, Wuhan University, Wuhan 430000 china}

\begin{abstract}
The next-to-leading order black-body radiation(BBR) shift to atomic energy-levels, namely $ (Z\alpha)^2\alpha T^2/m $ correction, was studied by using the nonrelativistic quantum electrodynamics(NRQED). We also estimate the one-loop contribution of quadrupole and the two-loop contributions of BBR-shift of the thermal(real) photon. These corrections have not been investigated before. The order of magnitude BBR-shift indicates the one-loop contribution of quadrupole is stronger than the previous result. And the two-loop contribution of BBR-shift of the thermal(real) photon is tiny, but this next-to-leading order BBR-shift may be as significant as the leading order in the multi-electron atoms or cold ones.  
\end{abstract}
\maketitle 

\section{INTRODCUTION}  
The blackbody radiation(BBR) can perturb the atomic energy-levels and shorten the lifetime of states. It is well known that the thermal mass-shift induced by BBR is the main part of the one-loop thermal self-energy correction\cite{Donoghue01}. The thermal mass-shift is proportional to $T^2$ and irrelevant to the atomic energy-level. The leading term of BBR-shift in the atom, which is proportional to $ T^4$, is the remainder term in the one-loop thermal self-energy correction. This effect is tiny. So it is always being neglected in the atomic energy-levels calculations. However, it is still necessary to consider BBR in some areas of atomic physics.  Firstly, in the precise calculation of the energy for simple atomic system, such as helium and lithium\cite{Schwartz00,Wang00}, the accuracy of  the energy of the ground state in free-field, has been extremely high, which means many higher-order corrections such as quantum electrodynamic(QED) effect, QED-nuclear recoil, QED-nuclear size, even if BBR-shift, could possibly be taken into account\cite{Eides00}. This is foremost to determine the physical fundamental constant, such as the fine-structure constant $\alpha$\cite{Pachucki00}.
Secondly, in these decades, the BBR effect becomes an obstacle of accuracy in determination of frequency standard\cite{Bloom00,Safronova00}. Several works\cite{Farley00,Itano00,Porsev00,Cooke00} have been devoted to study BBR-shift from 1980s. In those works, the classical BBR's electric field approximation and rotating-wave approximation were introduced. Escobedo\cite{Escobedo00} and Solovyev\cite{Solovyev00} studied the BBR-shift in the hydrogen-like atoms by using the finite temperature quantum electrodynamics approach without previous approximations. 
However, the BBR-shifts in multi-electron atoms, which the atomic clock is based on, haven't been studied by the same approach.  And those results\cite{Farley00,Itano00,Porsev00,Cooke00,Escobedo00,Solovyev00} are the leading order of the BBR-shift. It is beneficial to make a further step to study the higher corrections, for example, the $ (Z\alpha)^2\alpha T^2/m $ correction in this paper.

The bound state is cumbersome to be studied by using the quantum electrodynamics(QED). We present in this work an efficient way, nonrelativistic quantum electrodynamics(NRQED)\cite{Caswell00,Labelle00,Jentschura00,Pachucki01}, to attain the expected purposes. The NRQED is an effective field theory in which all virtual physics at scales greater than the electron mass have been integrated out. The advantage of using NRQED is that the perturbations can be calculated order by order. Imitating the finite temperature, QED\cite{Donoghue00,Donoghue01}, we replaced the photon propagator with the ensemble averaged photon propagator, when we adopted NRQED.  We've found a $ (Z\alpha)^2\alpha T^2/m $ correction, which is important and could be lager than the leading term of BBR-shift when the high/mid-$Z$ or cold hydrogen-like atoms are interrogated.  This $T^{2}$- dependent correction has not been investigated by former works yet. We conjecture this effect will be important in multi-electron atoms or cold ones. We also estimate the one-loop contribution of quadrupole and the two-loop contributions of BBR-shift of the thermal(real) photon. The order of magnitude BBR-shift indicates the one-loop contribution of quadrupole is stronger than the previous result\cite{•}. And the two-loop contribution of BBR-shift of the thermal(real) photon is tiny.
 In this paper, we studied BBR-shift by using NRQED approach and obtained the $ (Z\alpha)^2\alpha T^2/m $ correction in section 2. The  one-loop contribution of quadrupole and two-loop contributions of BBR-shift were estimated in section 3. The section 4 is devoted to discussion and conclusion.

\section{The $ (Z\alpha)^2\alpha T^2/m $ BBR-SHIFT TO THE ATOM}  
 The Feynman rules for the finite temperature QED are given in Fig.1, where $n_{B}(\omega)=\frac{1}{e^{\beta\omega}-1}$, $d_{ij}(\omega)=\left( \delta_{ij}-\frac{k_{i}k_{j}}{\omega^{2}}\right)$  and $\beta=\frac{1}{kT}$. 
\begin{figure}[htbp]
  \centering
  \includegraphics[width=4.0in,height=2.0in]{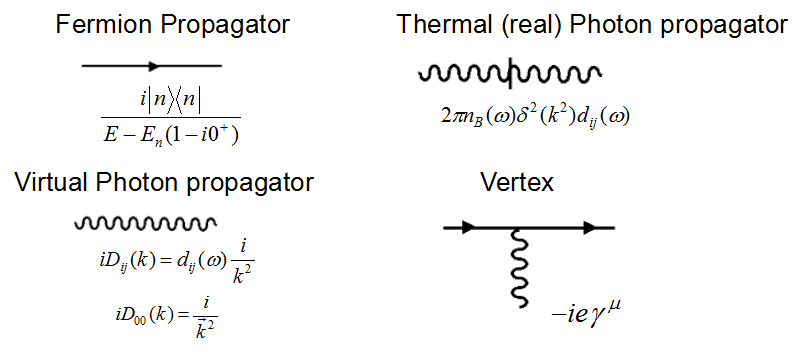}
  \caption{Feynman Rule}%\label{fig:digit}
\end{figure}
We adopt a many-body fermion propagator in Fig.1, where $\gamma^{\mu}$ are Dirac matrices and $|n\rangle,E_{n}$  are eigenfunction and eigenvalue of the many-body Hamiltonian respectively. Additionally, we adopted coulomb gauge to express the thermal(real) photon propagator, which is derived from the thermal(real) photon propagator\cite{Donoghue01} in Feynman gauge, by gauge transformation.    

The leading term of BBR-Shift is induced by the thermal one-loop self-energy correction, which is represented by diagrams in Fig.2. The contribution of virtual particle pairs[Fig.2(b)] is compressed by a $exp(-m/kT)\sim 0$ factor in the room temperature. So all the diagrams that have a thermal photon propagator connecting with virtual particle pairs can be neglected. The contribution of diagrams in Fig.2(a) is
 
\begin{figure}[htbp]
  \centering
  \includegraphics[width=3.5 in,height=1.2 in]{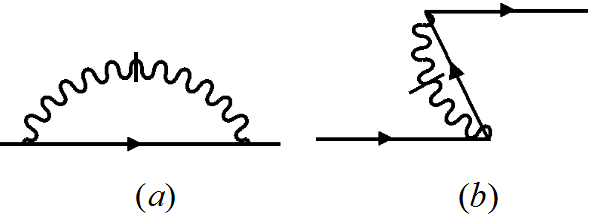}
  \caption{(a)Thermal one-loop self-energy correction of BBR-shift (Finite temperature QED). (b)The contribution of virtual particle pairs in hydrogen-like atoms, which is compressed by a $exp(-m/kT)\sim 0$ factor in the room temperature, can be neglected. }%\label{fig:digit}
\end{figure}

\begin{equation}
\begin{aligned}
\Delta E_{\mbox{2}} =&  e^{2} \mathcal{P}\int\frac{ n_{B}(\omega) d^{3}k}{(2\pi)^{3}2\omega} d_{ij}(\omega)
 \sum_{a,b}
\langle \psi|
\left( \gamma^{i} e^{-i \textbf{k}\cdot\textbf{x}}\right) _{a}
\dfrac{2(E-H)}{(E-H)^{2}-\omega^{2}}
\left( \gamma^{j} e^{i \textbf{k}\cdot \textbf{x}} \right) _{b}
|\psi\rangle,  \\
\end{aligned}
\end{equation}
where $\mathcal{P}$ denotes the Cauchy principal value, which is required for a proper treatment of resonant contributions\cite{Porsev00}.
It is coincided with Porsev's result\cite{Porsev00}, and contains bound-state-independent terms, which are called thermal mass-shift\cite{Donoghue01}. The remainder term is the BBR-shift. The thermal mass-shift also exists in free electron as well and is much larger than BBR-shift value. The contribution of the Fig.2 of the nonrelativistic approximation is represented as:

\begin{equation}
\begin{aligned}
\Delta E_{\textbf{2}}&=
\dfrac{4\alpha}{3\pi } \sum_{a,b} 
\mathcal{P}\int n_{B} \omega d\omega\langle \psi|
\left(\dfrac{\vec{p}}{m}\right)_{a}
\dfrac{2(E-H)}{(E-H)^{2}-\omega^{2}}
\left(\dfrac{\vec{p}}{m}\right)_{b}
|\psi\rangle+...
\\
& =-\dfrac{\pi\alpha}{3m\beta^{2}}N_{e}+\Delta E_{\textbf{2E1}}+... ,
\end{aligned}
\end{equation}
where the first term is the thermal mass-shift, the second 
term is the BBR-shift of electric dipole moment$(E1)$ and the ellipsis is the higher order term such as the contribution of multi-dipole. The $N_e$ is the number of the electrons, and $\alpha ( =e^{2}/(4\pi)\simeq10^{-4} ) $ is the fine-structure constant. The contribution of electric dipole$(E1)$ is
\begin{equation}
\begin{aligned}
\Delta E_{\textbf{2E1}}=
\dfrac{4\alpha}{3\pi } \sum_{a,b} 
\mathcal{P}\int n_{B}(\omega)\omega^{3} d\omega
\langle \psi|\left( \vec{r}\right)_{a} 
\dfrac{2(E-H)}{(E-H)^{2}-\omega^{2}}
\left( \vec{r}\right)_{b}|\psi\rangle
\end{aligned}
\end{equation}
where $\left( \vec{r}\right)_{a} $ and $\left( \vec{r}\right)_{b} $ are the position vectors of electron $a$ and $b$ respectively. 

 The BBR's characteristic wave-length $1//kT\simeq 10^{-5}m$ in the room temperature is much larger than the atoms' radius, and the electron is nonrelativistic in the light atoms. Additionally, the contributions of multi-polar moment are suppressed by $\alpha^{2}\simeq 10^{-4}$ (Magnetic dipole(M1)) or $(kT/m\alpha)^{2}\simeq 10^{-10}$ (Eletric-quadrupole(E2)) comparing with the contribution of the electric dipole(E1) in the room temperature\cite{Porsev00}. So the Eq.(3) is the leading term of BBR-shift. 
  
The Eq.(3) can be further simplified in the low-lying states in the room temperature(The approximation $E_{ab}\gg kT$ is applied to our work). Thus we have  
\begin{equation} 
\Delta E_{\textbf{2}}\simeq\frac{4\alpha\pi^{3}}{45\beta^{4}}\sum_{E_{\psi}\neq E_{M}}\frac{|\langle M|\vec{r}|\psi\rangle|^{2}}{E_{\psi M}},
 \end{equation}
which is proportional to $ T^4$. 
  
  The relativistic corrections aren't included in Eq.(3), and these corrections can be introduced by the nonrelativistic QED approach\cite{Jentschura00} 
\begin{equation}
\begin{aligned}
\Delta E_{\textbf{2R}} = & \dfrac{-4e^{2}}{3m_{a}m_{b}} 
\mathcal{P}\int\frac{n_{B}(\omega)d^{3}k}{(2\pi)^{3}2\omega} 
d_{ij}(\omega) \langle \psi|
 \sum_{a,b}
\Bigg\{
\left( \textbf{P}^{i} \right) _{a}
\dfrac{1}{E-H-\omega}
(V -\langle V\rangle)
\dfrac{1}{E-H-\omega}
 \left(\textbf{P}^{j} \right) _{b}
\\&+
2V \dfrac{1}{E-H}
\left( \textbf{P}^{i} \right) _{a}
\dfrac{1}{E-H-\omega}+(\omega\rightarrow -\omega)\Bigg\}
|\psi\rangle,
\end{aligned}
\end{equation}
where the inserted vertex $iV$ is relativistic correction and Breit interaction,

\begin{equation}
\begin{aligned}
V=& \sum_{a}\dfrac{-\textbf{P}^{4}_{a}}{8m^{3}} 
+ \sum_{a,b} \Bigg\{ -q_{a}q_{b}
\left( \dfrac{1}{8m_{a}^{2}}+\dfrac{1}{8m_{a}^{2}}\right) 
\delta^{3}(\textbf{r}_{ab}) 
\\&
-\dfrac{q_{a}q_{b}}{2m_{a}m_{b}}\dfrac{1}{4\pi r_{ab}}
\left( \textbf{P}_{a}\cdot\textbf{P}_{b}+
\dfrac{\textbf{r}_{ab}\cdot(\textbf{r}_{ab}\cdot\textbf{P}
_{b})\textbf{P}_{a}}{r_{ab}^{2}}\right)    \\
&-\dfrac{q_{a}q_{b}}{2m_{a}m_{b}} 
\dfrac{
\textbf{r}_{ab}\times\textbf{P}_{a}\cdot\boldsymbol{\sigma}
_{b}
+
\textbf{r}_{ba}\times\textbf{P}_{b}\cdot\boldsymbol{\sigma}
_{a}
}{4\pi r_{ab}^{3}}                           
-\dfrac{q_{a}q_{b}}{4m_{a}^{2}} 
\dfrac{
\textbf{r}_{ab}\times\textbf{P}_{a}\cdot\boldsymbol{\sigma}
_{a}}{4\pi r_{ab}^{3}} 
-\dfrac{q_{a}q_{b}}{4m_{b}^{2}} 
\dfrac{
\textbf{r}_{ba}\times\textbf{P}_{b}\cdot\boldsymbol{\sigma}
_{b}}{4\pi r_{ab}^{3}}      \\
&+\dfrac{q_{a}q_{b}}{2m_{a}m_{b}} 
\left[ \dfrac{
\boldsymbol{\sigma}_{a}\cdot\boldsymbol{\sigma}_{b}}{r_{ab}
^{3}}
+\dfrac{
\boldsymbol{3\sigma}_{a}\cdot\boldsymbol{r}_{ab}
\boldsymbol{\sigma}_{b}\cdot\boldsymbol{r}_{ab}}
{r_{ab}^{5}}
-\dfrac{2\boldsymbol{\sigma}_{a}\cdot\boldsymbol{\sigma}_{b}
\delta^{3}(\textbf{r}_{ab})}{3}
\right]\Bigg\}. 
\end{aligned}
\end{equation}
If $E_{ab}\gg kT$, the correction Eq.(5) can be written as
\begin{equation}
\begin{aligned}
\Delta E_{\textbf{2R}} = & -\dfrac{8\alpha\pi}{27\beta^{2}} \sum_{M_{1,2}}
\big(
\langle \psi|\left( \vec{r}\right)_{a} |M_{2}\rangle \cdot
\langle M_{1}|\left( \vec{r}\right)_{b}|\psi\rangle
\langle M_{2} |(V- \langle V\rangle )|M_{1}\rangle+
\\&
\langle \psi| V |M_{2}\rangle 
\langle M_{2} | \left( \vec{r}\right)_{a}|M_{1}\rangle
\cdot
\langle M_{1}|\left( \vec{r}\right)_{b}|\psi\rangle
\dfrac{E_{M_{1}M_{2}}}{E_{\psi M_{2}}}-
\\&
\langle \psi|\left( \vec{r}\right)_{a} |M_{2}\rangle 
\cdot
\langle M_{2} | \left( \vec{r}\right)_{b} |M_{1}\rangle
\langle M_{1}|V|\psi\rangle
\dfrac{E_{M_{1}M_{2}}}{E_{\psi M_{1}}}
\big),
\end{aligned}
\end{equation}
which is proportional to $ T^2$ Its magnitude is $(Z\alpha)^2\alpha T^2/m$.

The relativistic correction to current induces another correction\cite{Jentschura00}
\begin{equation}
\begin{aligned}
\Delta E_{\textbf{2J}} =&  2 e^{2} \mathcal{P}\int\frac{ n_{B}(\omega) d^{3}k}{(2\pi)^{3}2\omega} d_{ij}(\omega)
 \sum_{a,b}
\langle \psi|
\left( \delta j^{i}\right) _{a}
\dfrac{2(E-H)}{(E-H)^{2}-\omega^{2}}
\left(\dfrac{p^{j}}{m}\right) _{b}
|\psi\rangle,  \\
\end{aligned}
\end{equation}
where the current is
\begin{equation}
\begin{aligned}
(\delta j^{i})_{a}=\dfrac{-p^{i}\textbf{p}^{2}}{2m^{3}}+\sum_{c\neq a}\dfrac{(\alpha q_{c}\textbf{r}_{ac}\times\sigma)^{i}}{4 m^{2} r_{ac}^{3}}+\dfrac{i\omega(\textbf{p}\times\sigma)^{i}}{4m^{2}}+
\dfrac{(\textbf{k}\times\sigma)^{i}(\textbf{k}\cdot\textbf{r})}{2m},
\end{aligned}
\end{equation}
where $q_{c}$is the charge number of the particle $c$. The first two terms are proportional to $T^2$, the third term is proportional to $T^3$ and the last term is proportional to $T^4$ in the low-lying atomic states in the room temperature.

The Eq.(5)(8) are the $ (Z\alpha)^2\alpha T^2/m $ corrections.
These corrections have not been investigated before.
\section{THE QUADRUPOLE'S CONTRIBUTION AND TWO-LOOP CONTRIBUTION OF BBR-SHIFT}
It is obvious that Eq.(6) doesn't contain the contribution of multipole of the thermal photon, which can be obtained from Eq.(1). For example, The quadratic term is 
 
\begin{equation}
\begin{aligned}
\Delta E_{\mbox{2Q}} =& \dfrac{ e^{2}}{ m^{2}} \mathcal{P}\int\frac{ n_{B}(\omega) d^{3}k}{(2\pi)^{3}2\omega} d_{ij}(\omega)
 \sum_{a,b}
\langle \psi|\bigg[
 \left\lbrace \textbf{p}^{i} ,(-i \textbf{k}\cdot\textbf{x})\right\rbrace_{a}
\dfrac{2(E-H)}{(E-H)^{2}-\omega^{2}}
\left\lbrace \textbf{p}^{j} , (i \textbf{k}\cdot\textbf{x}) \right\rbrace  _{b}
\\&
+\left\lbrace  \textbf{p}^{i}, (i \textbf{k}\cdot\textbf{x})^{2}\right\rbrace _{a}
\dfrac{2(E-H)}{(E-H)^{2}-\omega^{2}}
\left(  \textbf{p}^{j} \right)_{b}\bigg]
|\psi\rangle  \\
=&  \dfrac{ e^{2}}{ m^{2}} \mathcal{P}\int\frac{ n_{B}(\omega)\omega^{3} d\omega}{(2\pi)^{2}} 
\dfrac{4\delta_{ij}\delta_{kl}-\delta_{ik}\delta_{jl}
-\delta_{il}\delta_{jk}}{15}
 \sum_{a,b}
\langle \psi|\bigg[
\left\lbrace  \textbf{p}^{i},\textbf{x}^{k}\right\rbrace_{a}
\dfrac{2(E-H)}{(E-H)^{2}-\omega^{2}}
\left\lbrace  \textbf{p}^{j},\textbf{x}^{l}\right\rbrace _{b}
\\&
-\left\lbrace \textbf{p}^{i},\textbf{x}^{k} \textbf{x}^{l} \right\rbrace _{a}
\dfrac{2(E-H)}{(E-H)^{2}-\omega^{2}}
\left(  \textbf{p}^{j} \right)_{b}\bigg]
|\psi\rangle.  \\
\end{aligned}
\end{equation}
If $E_{ab}\gg kT$, this correction of quadrupole is proportional to $T^4$. This correction is also different from the result in ref\cite{Porsev00}, which is proportional to $T^6$.
   
In this section, we will study the two-loop contributions of BBR-shift, which can be obtained by add a virtual or thermal photon propagator to the Fig.2(a).

\begin{figure}[htbp]
  \centering
  \includegraphics[width=5.0in,height=1.0 in]{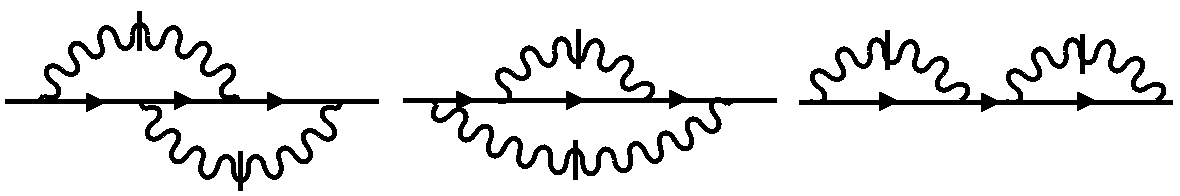}
  \caption{The thermal two-loop correction(Finite temperature QED). }%\label{fig:digit}
\end{figure}
 
\begin{figure}[htbp]
  \centering
  \includegraphics[width=5in,height=3in]{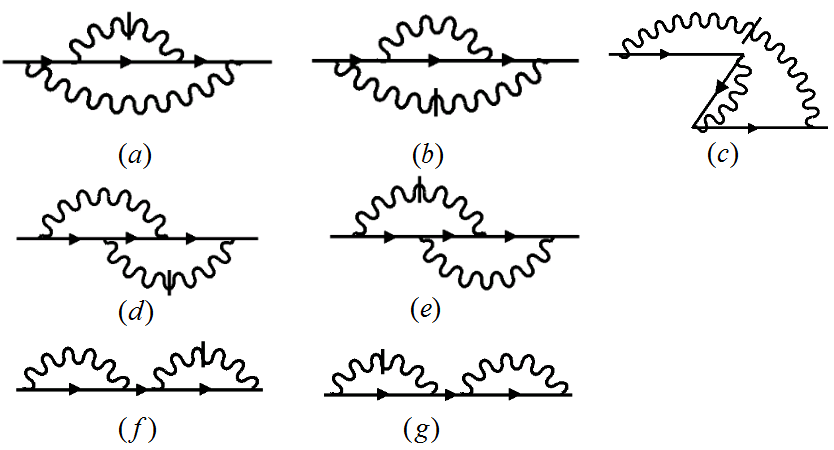}
  \caption{The mixing two-loop correction(Finite temperature QED). }%\label{fig:digit}
\end{figure}

 The Fig.3 are thermal two-loop diagrams in multi-electron atoms(Both the photon propagators are thermal(real) propagators). Its contribution is
\begin{equation}
\begin{aligned}
\Delta E_{\textbf{3}} & =  e^{4}
\sum_{n_{1}n_{2}n_{3}}
\sum_{a,b,c,d}
 \mathcal{P}\int\frac{n_{B}(\omega)d^{3}k}{(2\pi)^{3}2\omega}
 \frac{n_{B}(\omega')d^{3}k'}{(2\pi)^{3}2\omega'}
 (\delta_{il}-\frac{k_{i}k_{l}}{\omega^{2}})   
 (\delta_{jk}-\frac{k'_{j}k'_{k}}{\omega'^{2}})        
\\& 
\langle\psi|\left(\gamma^{l} e^{i\vec{k}\cdot\vec{x}}\right)_{d} |n_{3}\rangle
\langle n_{3}|\left( \gamma^{k} e^{i\vec{k'}\cdot\vec{x}}\right)_{c} |n_{2}\rangle
\langle n_{2}|\left( \gamma^{j} e^{-i\vec{k'}\cdot\vec{x}}\right)_{b} |n_{1}\rangle
\langle n_{1}|\left( \gamma^{i} e^{-i\vec{k}\cdot\vec{x}}\right)_{a} |\psi\rangle 
\\& 
\bigg[
\frac{1}{E_{\psi n_{3}}-\omega} 
\frac{2(E_{\psi n_{2}}-\omega)}{(E_{\psi n_{2}}-\omega)^{2}-\omega'^{2}} 
\frac{1}{E_{\psi n_{1}}-\omega}+
\frac{1}{E_{\psi n_{3}}+\omega} 
\frac{2(E_{\psi n_{2}}+\omega)}{(E_{\psi n_{2}}+\omega)^{2}-\omega'^{2}} 
\frac{1}{E_{\psi n_{1}}+\omega}\bigg]+                 
\\&
 e^{4}
\sum_{n_{1}n_{2}n_{3}}
\sum_{a,b,c,d}
\int
\int\frac{n_{B}(\omega)d^{3}k}{(2\pi)^{3}2\omega}
 \frac{n_{B}(\omega')d^{3}k'}{(2\pi)^{3}2\omega'}
(\delta_{ik}-\frac{k_{i}k_{k}}{\omega^{2}})
(\delta_{jl}-\frac{k'_{j}k'_{l}}{\omega'^{2}})   
\\&
\langle \psi|\left( \gamma^{l} e^{i\vec{k'}\cdot\vec{x}}\right)_{d} |n_{3}\rangle
\langle n_{3}|\left( \gamma^{k} e^{i\vec{k}\cdot\vec{x}}\right)_{c} |n_{2}\rangle
\langle n_{2}|\left( \gamma^{j} e^{-i\vec{k'}\cdot\vec{x}}\right)_{b} |n_{1}\rangle
\langle n_{1}|\left( \gamma^{i} e^{-i\vec{k}\cdot\vec{x}}\right)_{a} |\psi\rangle \\
& \bigg[\frac{1}{E_{\psi n_{3}}-\omega'} \frac{2(E_{\psi n_{2}}-\omega')}{(E_{\psi n_{2}}-\omega')^{2}-\omega^{2}} \frac{1}{E_{\psi n_{1}}-\omega}+
\frac{1}{E_{\psi n_{3}}+\omega'} \frac{2(E_{\psi n_{2}}+\omega')}{(E_{\psi n_{2}}+\omega')^{2}-\omega^{2}} \frac{1}{E_{\psi n_{1}}+\omega}\bigg]+                 
\\&
 e^{4}
\sum_{n_{1}n_{2}n_{3}}
\sum_{a,b,c,d}
\int
\int\frac{n_{B}(\omega)d^{3}k}{(2\pi)^{3}2\omega}
 \frac{n_{B}(\omega')d^{3}k'}{(2\pi)^{3}2\omega'}
(\delta_{ij}-\frac{k_{i}k_{j}}{\omega^{2}})
(\delta_{kl}-\frac{k'_{k}k'_{l}}{\omega'^{2}})   
\\&
\langle \psi|\left( \gamma^{l} e^{i\vec{k'}\cdot\vec{x}}\right)_{d} |n_{3}\rangle
\langle n_{3}|\left( \gamma^{k} e^{-i\vec{k'}\cdot\vec{x}}\right)_{c} |n_{2}\rangle
\langle n_{2}|\left( \gamma^{j} e^{i\vec{k}\cdot\vec{x}}\right)_{b} |n_{1}\rangle
\langle n_{1}|\left( \gamma^{i} e^{-i\vec{k}\cdot\vec{x}}\right)_{a} |\psi\rangle \\
& \frac{2E_{\psi n_{3}}}{E_{\psi n_{3}}^{2}-\omega'^{2}} \frac{1}{E_{\psi n_{2}}-\omega} \frac{2E_{\psi n_{1}}}{E_{\psi n_{1}}^{2}-\omega^{2}},
\end{aligned}
\end{equation}
where $E_{\psi M}=E_{\psi}(1-i0^{+})-E_{M}(1-i0^{+})$ is the difference between energys. The contribution of the electric dipole of Eq.(11) is the more important then multipoles' in the room temperature, which can be obtained by $\gamma^{i}\rightarrow P^{i}/m$. This integral is finite both in ultraviolet and infrared regions. 

Other diagrams two-loop contributions are the mixing two-loop diagram(Fig.4). The Fig.4(a) is the easiest to 
be obtained,

\begin{equation}
\begin{aligned}
\Delta E_{\textbf{4a}} & =  e^{4}
\sum_{n_{1}n_{2}n_{3}}
\sum_{a,b,c,d}
 \mathcal{P}\int\frac{d^{3}k}{(2\pi)^{3}2\omega}
 \frac{n_{B}(\omega')d^{3}k'}{(2\pi)^{3}2\omega'}
 (\delta_{il}-\frac{k_{i}k_{l}}{\omega^{2}})   
 (\delta_{jk}-\frac{k'_{j}k'_{k}}{\omega'^{2}})        
\\& 
\langle \psi|\left(\gamma^{l} e^{i\vec{k}\cdot\vec{x}}\right)_{d} |n_{3}\rangle
\langle n_{3}|\left(\gamma^{k} e^{i\vec{k'}\cdot\vec{x}}\right)_{c} |n_{2}\rangle
\langle n_{2}|\left(\gamma^{j} e^{-i\vec{k'}\cdot\vec{x}}\right)_{b} |n_{1}\rangle
\langle n_{1}|\left(\gamma^{i} e^{-i\vec{k}\cdot\vec{x}}\right)_{a} |\psi\rangle 
\\& 
\frac{1}{E_{\psi n_{3}}-\omega} 
\frac{2(E_{\psi n_{2}}-\omega)}{(E_{\psi n_{2}}-\omega)^{2}-\omega'^{2}} 
\frac{1}{E_{\psi n_{1}}-\omega}.               
\end{aligned}
\end{equation}
This integral is also finite. The remainder parts in Fig.4 are divergence. These integrals must subtract a counterterm, and the differences are their contribution of BBR-shift. We haven't obtained this difference. However, Its order of magnitude can be easily estimated(We will list in the next section). They are $\alpha$ weaker than (5)(8).

\section{DISCUSSION AND CONCLUSION}
We have studied the two-loop corrections of BBR-shift in section.3. Although some operators of corrections haven't been obtained, we could estimate their order of magnitude in the light Hydrogen-like atoms. Attribute to the order counting rules of correction terms $\langle p\rangle\sim1/\langle r\rangle\sim mZ\alpha$,$\langle E\rangle \sim m(Z\alpha)^{2} $ in the light Hydrogen-like atoms, The dimension parts of the BBR-shift of low-lying states are listed in Table.I. Two approximations have been applied, which are nonrelativistic and electric-dipole approximation, and the energy-gaps between low-lying states satisfying $\Delta E \gg kT$.     

\begin{table}[htbp]
\caption{The magnitude of BBR-shift$(Hz)$. They are listed by the descending order of $\alpha $ factor. 
$\langle\Delta E_{\textbf{2Ji}}\rangle$ is energy-shift Eq.(8) originating from the $i$th term in Eq.(9). Two approximations have been applied, such as (1)nonrelativistic electron and electric-dipole approximation of the thermal photon. (2) The energy-gaps between low-lying states $\Delta E \gg kT$.}
\begin{tabular}{cc}
     \hline\hline
     &The magnitude of BBR-shift\\
     \hline
     $\delta m(T)=-\frac{\alpha\pi}{3m\beta^{2}}$& $2.42\times10^{3}\frac{(T)^{2}}{300^{2}}$
     \\
     $\langle\Delta E_{\textbf{2E1}}\rangle
     \sim\frac{1}{Z^{4}m^{3}\alpha^{3}\beta^{4}} $ &
     $\frac{10^{-3}}{Z^{4}}\frac{T^{4}}{300^{4}}$ 
     \\
     $\langle\Delta E_{\textbf{2Q}}\rangle=
     \langle\Delta E_{\textbf{2J4}}\rangle
     \sim\frac{1}{Z^{2}m^{3}\alpha\beta^{4}} $ &
     $\frac{10^{-7}}{Z^{2}}\frac{T^{4}}{300^{4}}$ 
     \\
      $\langle\Delta
     E_{\textbf{3}}\rangle\sim\frac{1}{m^{3}Z^{2}\beta^{4}}$&$\frac{10^{-9}}{Z^{2}}\frac{T^{4}}{300^{4}}$
     \\
     $\langle\Delta E_{\textbf{2J3}}\rangle
     \sim\frac{\alpha}{m^{2}\beta^{3}} $ &
     $10^{-4}\frac{T^{3}}{300^{3}}$ 
     \\
     $\langle\Delta E_{\textbf{2J2}}\rangle
     \sim\frac{(Z\alpha)\alpha^{2}}{m\beta^{2}}
      $ &
      $10^{-1}\frac{ZT^{2}}{300^{2}}$
     \\
     $\langle\Delta E_{\textbf{2R}}\rangle
     =\langle\Delta E_{\textbf{2J1}}\rangle
     \sim\frac{(Z\alpha)^{2}\alpha}{m\beta^{2}}$&$10^{-1}\frac{(ZT)^{2}}{300^{2}}$
     \\    
  $\langle\Delta E_{\textbf{4}}\rangle\sim\frac{(Z\alpha)^{2}\alpha^{2}}{m\beta^{2}}$&$10^{-3}\frac{(ZT)^{2}}{300^{2}}$\\
     \hline
\end{tabular}
\end{table}

The results in table.I are arranged by the descending order of 
$\alpha$ factor. One interesting thing is the magnitude of the contribution of quadrupole is proportional to $T^4$. This result is stronger than the previous result\cite{Porsev00}, which is suppressed by a $(kT/m)^2=10^{-6}$ factor in the room temperature. The two-loop correction of the thermal photon is so tiny comparing with other corrections. The Eq.(3) ($\langle\Delta E_{\textbf{2E1}}\rangle$) is the leading term of BBR-shift, although the order of magnitude of Eq.(5)(8)  ($\langle\Delta E_{\textbf{2R}}\rangle,\langle\Delta E_{\textbf{2J1}}\rangle$) is larger than the Eq.(3) in the Table.I. What most significant is that comparing with leading term, when charge $Z$ is higher or $T$ is lower, the  $(Z\alpha)^2\alpha T^2/m $ correction will become more important(Because $\langle\Delta E_{\textbf{3}}\rangle/\langle\Delta E_{\textbf{2}}\rangle\sim Z^{6}/T^{2} $). This correction  $(Z\alpha)^2\alpha T^2/m $ must be recalculated seriously in the future. We also conjecture that it may be important in the multi-electron atoms or cold ones.   

The reasons (1) why the $\langle\Delta E_{\textbf{2R}}\rangle
,\langle\Delta E_{\textbf{2J1}}\rangle$ are significant,
(2)why the magnitude of the contribution of quadrupole is proportional to $T^4$ rather than $T^6$ in ref\cite{Porsev00}. 
(3)why $\langle\Delta E_{\textbf{4}}\rangle$ is proportional to $T^4$ are the same: These BBR-shift is mixing thermal mass-shift($\varpropto T^2$) with counterpart corrections.

\begin{acknowledgments}
This work was supported by the National Natural Science Foundation of China(No.11274246) 
\end{acknowledgments}


\begin{thebibliography}{35}
\expandafter\ifx\csname
natexlab\endcsname\relax\def\natexlab#1{#1}\fi
\expandafter\ifx\csname bibnamefont\endcsname\relax
  \def\bibnamefont#1{#1}\fi
\expandafter\ifx\csname bibfnamefont\endcsname\relax
  \def\bibfnamefont#1{#1}\fi
\expandafter\ifx\csname citenamefont\endcsname\relax
  \def\citenamefont#1{#1}\fi
\expandafter\ifx\csname url\endcsname\relax
  \def\url#1{\texttt{#1}}\fi
\expandafter\ifx\csname
urlprefix\endcsname\relax\def\urlprefix{URL }\fi
\providecommand{\bibinfo}[2]{#2}
\providecommand{\eprint}[2][]{\url{#2}}


\bibitem[{\citenamefont{Donoghue et~al.}(1985)\citenamefont{Donoghue, Holstein and Robinett}}]{Donoghue01}
\bibinfo{author}{\bibfnamefont{J.~F }\bibnamefont{Donoghue}},
\bibinfo{author}{\bibfnamefont{B.~R.}\bibnamefont{Holstein}}
\bibnamefont{and} \bibinfo{author}{\bibfnamefont{R.~W.}\bibnamefont{Robinett}},
\bibinfo{journal}{Ann.Phys.(N.Y.).}
\textbf{\bibinfo{volume}{164}},\bibinfo{pages}{233}
(\bibinfo{year}{1985}).



\bibitem[{\citenamefont{Schwartz.}(2006)\citenamefont{Schwartz}}]{Schwartz00}
\bibinfo{author}{\bibfnamefont{C. }~\bibnamefont{Schwartz}},
\bibinfo{journal}{Int. J. Mod. Phy. E.}
\textbf{\bibinfo{volume}{15}}, \bibinfo{nonumber}{4},\bibinfo{pages}{877}
(\bibinfo{year}{2006}).


\bibitem[{\citenamefont{Wang et~al.}(2011)\citenamefont{Wang,Yan,Qiao and Drake}}]{Wang00}
\bibinfo{author}{\bibfnamefont{L.~M. }\bibnamefont{Wang}},
\bibinfo{author}{\bibfnamefont{Z.~C.} \bibnamefont{Yan}},
\bibinfo{author}{\bibfnamefont{H.~X.} \bibnamefont{Qiao}}
\bibnamefont{and} \bibinfo{author}{\bibfnamefont{G.~W.~F.}\bibnamefont{Drake}},
\bibinfo{journal}{Phys. Rev. A.}
\textbf{\bibinfo{volume}{83}}, \bibinfo{pages}{034503}(2011)

\bibitem[{\citenamefont{Eides et~al.}(2001)\citenamefont{Eides, Grotch and Shelyuto}}]{Eides00}
\bibinfo{author}{\bibfnamefont{M.~I. }\bibnamefont{Eides}},
\bibinfo{author}{\bibfnamefont{H.}~ \bibnamefont{Grotch}},
\bibnamefont{and} \bibinfo{author}{\bibfnamefont{V.~A.}\bibnamefont{Shelyuto}},
\bibinfo{journal}{Physics.Reports.}
\textbf{\bibinfo{volume}{342}}, \bibinfo{Issues}{2-3}, \bibinfo{pages}{63}(2001)


\bibitem[{\citenamefont{Pachucki et~al.}(2010)\citenamefont{Pachucki and Yerokhin}}]{Pachucki00}
\bibinfo{author}{\bibfnamefont{K. }~\bibnamefont{Pachucki}}
\bibnamefont{and} \bibinfo{author}{\bibfnamefont{Y. ~A.}\bibnamefont{Yerokhin}},
\bibinfo{journal}{Phys. Rev. Lett.}
\textbf{\bibinfo{volume}{104}},\bibinfo{pages}{070403}
(\bibinfo{year}{2010}).



\bibitem[{\citenamefont{Bloom et~al.}(2014)\citenamefont{Bloom,Nicholson,Williams,Campbell, Bishof,Zhang,Zhang,Bromley, and Ye}}]{Bloom00}
\bibinfo{author}{\bibfnamefont{B.~J.}\bibnamefont{Bloom}},
\bibinfo{author}{\bibfnamefont{T.~L.}\bibnamefont{Nicholson}},
\bibinfo{author}{\bibfnamefont{J.~R.}\bibnamefont{Williams}},
\bibinfo{author}{\bibfnamefont{S.~L.}\bibnamefont{Campbell}},
\bibinfo{author}{\bibfnamefont{M.}~\bibnamefont{Bishof}},
\bibinfo{author}{\bibfnamefont{X.}~\bibnamefont{Zhang}},
\bibinfo{author}{\bibfnamefont{W.}~\bibnamefont{Zhang}},
\bibinfo{author}{\bibfnamefont{S.~L.}\bibnamefont{Bromley}},
\bibnamefont{and} \bibinfo{author}{\bibfnamefont{J.}~\bibnamefont{Ye}},
\bibinfo{journal}{Nature(London).}
\textbf{\bibinfo{volume}{506}}, \bibinfo{pages}{71}
(\bibinfo{year}{2014}).


\bibitem[{\citenamefont{Safronova et~al.}(2013)\citenamefont{Safronova,Porsev,Safronova,Kozlovl, and Clark}}]{Safronova00}
\bibinfo{author}{\bibfnamefont{M. ~S.}\bibnamefont{Safronova}},
\bibinfo{author}{\bibfnamefont{S. ~G.} \bibnamefont{Porsev}},
\bibinfo{author}{\bibfnamefont{U. ~I.} \bibnamefont{Safronova}},
\bibinfo{author}{\bibfnamefont{M. ~G.}\bibnamefont{Kozlov}},
\bibnamefont{and} \bibinfo{author}{\bibfnamefont{C.~W.}\bibnamefont{Clark}},
\bibinfo{journal}{Phys. Rev. A.}
\textbf{\bibinfo{volume}{87}}, \bibinfo{pages}{012509}
(\bibinfo{year}{2013}).


\bibitem[{\citenamefont{Farley et~al.}(1981)\citenamefont{Farley and Wing}}]{Farley00}
\bibinfo{author}{\bibfnamefont{J.~W.} \bibnamefont{Farley}}
\bibnamefont{and} \bibinfo{author}{\bibfnamefont{W.~H.}\bibnamefont{Wing}},
\bibinfo{journal}{Phys. Rev. A.}
\textbf{\bibinfo{volume}{23}}, \bibinfo{pages}{2397}
(\bibinfo{year}{1981}).

\bibitem[{\citenamefont{Itano et~al.}(1981)\citenamefont{Itano, Lewis and Wineland}}]{Itano00}
\bibinfo{author}{\bibfnamefont{W.~M.} \bibnamefont{Itano}},
\bibinfo{author}{\bibfnamefont{L.~L.} \bibnamefont{Lewis}},
\bibnamefont{and} \bibinfo{author}{\bibfnamefont{D.~J.}\bibnamefont{Wineland}},
\bibinfo{journal}{Phys. Rev. A.}
\textbf{\bibinfo{volume}{25}}, \bibinfo{pages}{1233}
(\bibinfo{year}{1982}).



\bibitem[{\citenamefont{Porsev et~al.}(2006)\citenamefont{Porsev and Derevianko}}]{Porsev00}
\bibinfo{author}{\bibfnamefont{G. }~\bibnamefont{Porsev}}
\bibnamefont{and} \bibinfo{author}{\bibfnamefont{A.}~\bibnamefont{Derevianko}},
\bibinfo{journal}{Phys. Rev. A.}
\textbf{\bibinfo{volume}{74}}, \bibinfo{pages}{020502}
(\bibinfo{year}{2006}).

\bibitem[{\citenamefont{Cooke et~al.}(1980)\citenamefont{Cooke and Gallagher}}]{Cooke00}
\bibinfo{author}{\bibfnamefont{W. E.}~\bibnamefont{Cooke}}
\bibnamefont{and} \bibinfo{author}{\bibfnamefont{T. ~F.}~\bibnamefont{Gallagher}},
\bibinfo{journal}{Phys. Rev. A.}
\textbf{\bibinfo{volume}{21}}, \bibinfo{pages}{588}
(\bibinfo{year}{1980}).



\bibitem[{\citenamefont{Escobedo et~al.}(2015)\citenamefont{Escobedo and Soto}}]{Escobedo00}
\bibinfo{author}{\bibfnamefont{M.~A }~\bibnamefont{Escobedo}},
\bibnamefont{and} \bibinfo{author}{\bibfnamefont{J.}~\bibnamefont{Soto}},
\bibinfo{journal}{Phys. Rev. A.}
\textbf{\bibinfo{volume}{78}}, \bibinfo{pages}{032520}
(\bibinfo{year}{2008}).



\bibitem[{\citenamefont{Solovyev et~al.}(2015)\citenamefont{Solovyev,Labzowsky, and Plunien}}]{Solovyev00}
\bibinfo{author}{\bibfnamefont{D. }~\bibnamefont{Solovyev}},
\bibinfo{author}{\bibfnamefont{L.}~ \bibnamefont{Labzowsky}},
\bibnamefont{and} \bibinfo{author}{\bibfnamefont{G.}~\bibnamefont{Plunien}},
\bibinfo{journal}{Phys. Rev. A.}
\textbf{\bibinfo{volume}{92}}, \bibinfo{pages}{022508}
(\bibinfo{year}{2015}).





\bibitem[{\citenamefont{Caswell et~al.}(1986)\citenamefont{Caswell and Plunien}}]{Caswell00}
\bibinfo{author}{\bibfnamefont{W.~E }\bibnamefont{Caswell}}
\bibnamefont{and} \bibinfo{author}{\bibfnamefont{G.~P}~\bibnamefont{Lepage}},
\bibinfo{journal}{Phys. Lett.}
\textbf{\bibinfo{volume}{167B}}, \bibinfo{number}{4},\bibinfo{pages}{437}
(\bibinfo{year}{1986}).


\bibitem[{\citenamefont{Labelle.}(1998)}]{Labelle00}
\bibinfo{author}{\bibfnamefont{P.}~\bibnamefont{Labelle}},
\bibinfo{journal}{Phys.Rev.D.}
\textbf{\bibinfo{volume}{58}},\bibinfo{pages}{093013}
(\bibinfo{year}{1998}).


\bibitem[{\citenamefont{Jentschura et~al.}(2005)\citenamefont{Jentschura, Czarnecki and  Pachucki}}]{Jentschura00}
\bibinfo{author}{\bibfnamefont{U.~D }\bibnamefont{Jentschura}},
\bibinfo{author}{\bibfnamefont{A. }\bibnamefont{Czarnecki}},
\bibnamefont{and} \bibinfo{author}{\bibfnamefont{K. }\bibnamefont{Pachucki}},
\bibinfo{journal}{Phys. Rev. A.}
\textbf{\bibinfo{volume}{72}},\bibinfo{pages}{062102}
(\bibinfo{year}{2005}).






\bibitem[{\citenamefont{Pachucki}(2005)
\citenamefont{Donoghue, and Holstein}}]{Pachucki01}
\bibinfo{author}{\bibfnamefont{K~ }\bibnamefont{Pachucki}},
\bibinfo{journal}{Phys. Rev. A.}
\textbf{\bibinfo{volume}{71}},\bibinfo{pages}{012503}
(\bibinfo{year}{2005}).






\bibitem[{\citenamefont{Donoghue et~al.}(1983)\citenamefont{Donoghue, and Holstein}}]{Donoghue00}
\bibinfo{author}{\bibfnamefont{J.~F }\bibnamefont{Donoghue}},
\bibnamefont{and} \bibinfo{author}{\bibfnamefont{B.~R.}\bibnamefont{Holstein}},
\bibinfo{journal}{Phys. Rev. D.}
\textbf{\bibinfo{volume}{28}},\bibinfo{pages}{340}
(\bibinfo{year}{1983}).



\bibitem[{\citenamefont{Lindgren.}(2011)\citenamefont{Lindgren}}]{Lindgren00}
\bibinfo{author}{\bibfnamefont{I. }~\bibnamefont{Lindgren}},
\bibinfo{book}{Relativistic.Many-Body.Theory.}
\bibinfo{pubulisher}( Springer , New York , 2011 ).
























\end{thebibliography}
\end{document}